\documentclass{article}
\usepackage[utf8]{inputenc}

\usepackage{authblk}
\usepackage{url}
\usepackage{graphicx}
\usepackage{fullpage}
\usepackage{xcolor}
\usepackage{booktabs}
\usepackage{caption}
\usepackage{subcaption}

\usepackage[capitalize]{cleveref}
\crefname{section}{Sec.}{Secs.}
\Crefname{section}{Section}{Sections}
\Crefname{table}{Table}{Tables}
\crefname{table}{Tab.}{Tabs.}

\title{COVID-Net UV: An End-to-End Spatio-Temporal Deep Neural Network Architecture for Automated Diagnosis of COVID-19 Infection from Ultrasound Videos}

\author[1]{Hilda Azimi\thanks{hilda.azimi@nrc-cnrc.gc.ca}}
\author[1]{Ashkan Ebadi\thanks{ashkan.ebadi@nrc-cnrc.gc.ca}}
\author[2]{Jessy Song\thanks{jessy.jia.song@uwaterloo.ca}}
\author[1]{Pengcheng Xi\thanks{pengcheng.xi@nrc-cnrc.gc.ca}}
\author[2]{Alexander Wong\thanks{alexander.wong@uwaterloo.ca }}
\affil[1]{National Research Council Canada}
\affil[2]{University of Waterloo}

\date{}

\begin{document}

\maketitle

\begin{abstract}
Besides vaccination, as an effective way to mitigate the further spread of COVID-19, fast and accurate screening of individuals to test for the disease is yet necessary to ensure public health safety. We propose COVID-Net UV, an end-to-end hybrid spatio-temporal deep neural network architecture, to detect COVID-19 infection from lung point-of-care ultrasound videos captured by convex transducers. COVID-Net UV comprises a convolutional neural network that extracts spatial features and a recurrent neural network that learns temporal dependence. After careful hyperparameter tuning, the network achieves an average accuracy of 94.44\% with no false-negative cases for COVID-19 cases. The goal with COVID-Net UV is to assist front-line clinicians in the fight against COVID-19 via accelerating the screening of lung point-of-care ultrasound videos and automatic detection of COVID-19 positive cases.
\end{abstract}

\section{Introduction}
The Coronavirus Disease 2019 (COVID-19) has resulted in a dramatic loss of life worldwide and posed an unprecedented public health challenge. There is no doubt that vaccination has been helping in mitigating the further spread of COVID-19. However, fast screening individuals to test for the disease is still necessary to ensure public health safety \cite{MacLean2021COVIDNet}. Chest x-ray (CXR) and computed tomography (CT) are two modalities that are often used for screening patients suspicious for COVID-19 infection. Another imaging modality for diagnosing lung-related diseases is the lung point-of-care ultrasound (POCUS). This modality has been suggested as the most helpful in contexts that are resource-limited, such as emergency settings or low-resource countries \cite{MacLean2021COVIDNet}. Compared to CXR and CT, POCUS is much cheaper to acquire and has higher portability and accessibility, thus enhancing the ability for possible COVID-19 screening \cite{Amatya2018Diagnostic}. Deep learning (DL) networks have been applied to POCUS images for different tasks and analyses such as segmentation, disease classification, and detection \cite{Liu2019Deep}. However, the protocol for physicians to perform an ultrasound (US) examination requires them to capture and analyze the US video, often from various angles, views, and positions \cite{Amatya2018Diagnostic,MacLean2021COVIDNet}. This means that the sequences of US video frames from one position or view to another can provide physicians with more information to make an accurate diagnosis; and perhaps not all frames of US videos contain signs and symptoms of a suspected disease. Therefore, applying DL to frames of US videos only and without considering their temporal information is not the ideal solution to adopt POCUS data for screening and diagnostics purposes. Motivated by this challenge, we propose COVID-Net UV, an end-to-end spatio-temporal deep neural network architecture to detect COVID-19 positive cases from POCUS videos. Our contributions can be summarized as follows: 1) COVID-Net UV is an effective tool for automatic detection of COVID-19 positive cases from POCUS videos without requiring the need for technician intervention and any further processing, 2) it also bridges the gap in the current diagnostic procedure of POCUS data by eliminating the need for the time-consuming and costly training of human experts, as interpreting US data requires domain knowledge \cite{Ng2020Imaging}.

\begin{figure}[t]
  \centering
   \includegraphics[width=0.99\linewidth]{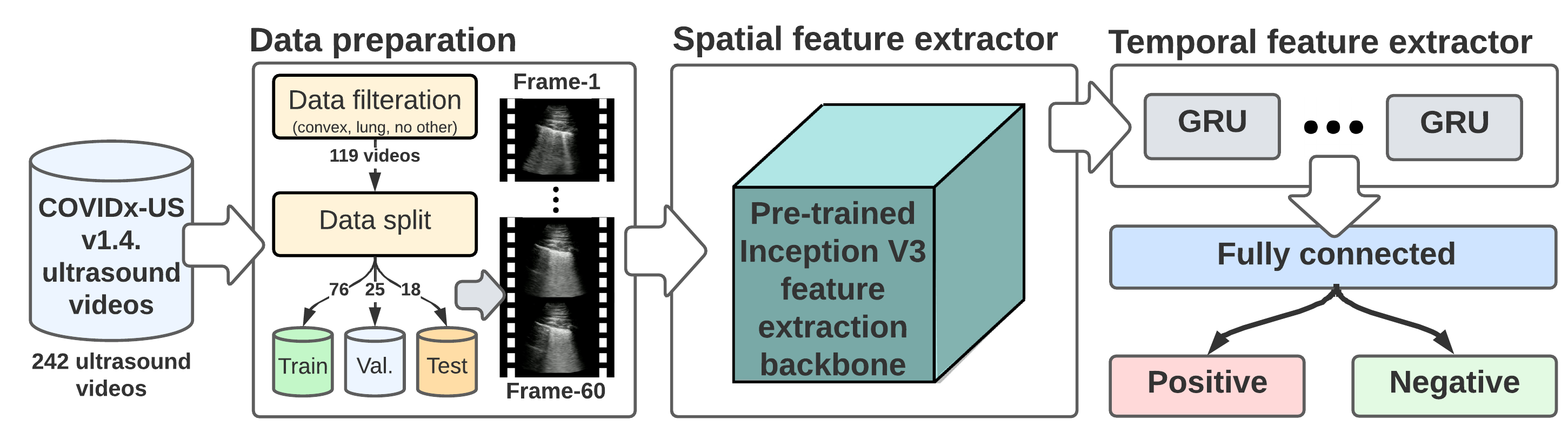}
   \caption{COVID-Net UV: a CNN-RNN architecture to classify POCUS videos in to two class of positive, i.e., COVID-19 infection, and negative, i.e., pneumonia or normal.}
   \label{fig:chart}
\end{figure}

\section{Related Work}
A number of techniques have been proposed so far for classification of various features in POCUS images and videos related explicitly to COVID-19 Disease, where some are models learned by DL, mostly using frame-based data as input. Roy et al. applied a DL model, derived from spatial transformer networks, to predict the COVID-19 severity score associated with POCUS videos frames, provide localization of pathological artifacts in a weakly-supervised way, and adopted a uninorms-based method for frame score aggregation at the video-level \cite{Roy2020Deep}. In \cite{Erfanian2021Automated}, authors presented a technique for classifying POCUS videos, based on a Two-Stream Inflated 3D ConvNet (I3D) to categorize the main imaging features seen in POCUS scans, such as Alines, B-lines, consolidation, and pleural effusion, which unveil the degree to which the lungs have been affected by the infection. Both these works had different approaches to analyzing PUCUS videos in the presence of COVID19 disease. We aim to bridge the gap in the current diagnostic procedure of POCUS Videos to detect and classify COVID-19 disease by learning spatio-temporal features, combining Convolutional Neural Network (CNN) and Recurrent Neural Network (RNN) architectures.

\section{Data and Methods}
To train and evaluate the COVID-Net UV, we used the COVIDx-US dataset v1.4. \cite{Ebadi2021COVIDx-US} that contains 242 US videos curated and integrated from 9 different data sources. The videos comprised four different classes: COVID-19 infection, non-COVID-19 infection, other lung diseases/conditions, and normal control cases. We filtered out the \textit{other} class, i.e., other lung diseases/conditions, due to the heterogeneity of the cases. And, we only included lung US video captured with a convex probe. We formulated the problem as a binary classification problem, i.e., the COVID-19 cases were labeled as positive and the normal and non-COVID-19 cases as negative. This resulted in 119 videos in total that were split into a training set with 76 videos (38 positives and 38 negatives), a validation set with 25 videos (12 positives and 13 negatives), and a test set with 18 videos (10 positives and 8 negatives).

We employed a hybrid architecture that included convolutional and recurrent layers to process spatial and temporal aspects of videos, respectively (\cref{fig:chart}). We adopted the InceptionV3 model pre-trained on the ImageNet data set \cite{Azimi2022Improving}, as the spatial feature extraction backbone and added 2 GRU units to capture temporal features. Since a video is an ordered sequence of frames, the frames can be extracted and placed on a 3D tensor. However, the number of frames may vary from video to video, making it impossible to stack them in batches. To overcome this problem, we first captured the frames of a video. Next, we extracted frames from the videos until a maximum frame count was reached. In this case, if the frame count was lower than the maximum frame count, the video was padded with zeros. We chose 60 as the maximum frame count, considering the characteristics of the videos in the dataset. Our training strategy involved two callbacks, i.e., \textit{learning rate scheduler} (decay learning rate by factor of 0.5 after three epochs with no performance improvement on validation set), and \textit{early stopping} (stop training after seven epochs with no performance improvement on validation set), to optimize our network and avoid over-fitting. The initial learning rate and the maximum number of epochs were set at 0.001 and 30, respectively. The network was trained for 18 epochs following the \textit{early stopping} strategy.

\section{Results}
\begin{figure}[t]
     \centering
     \begin{subfigure}[b]{0.31\textwidth}
         \centering
         \includegraphics[width=\textwidth]{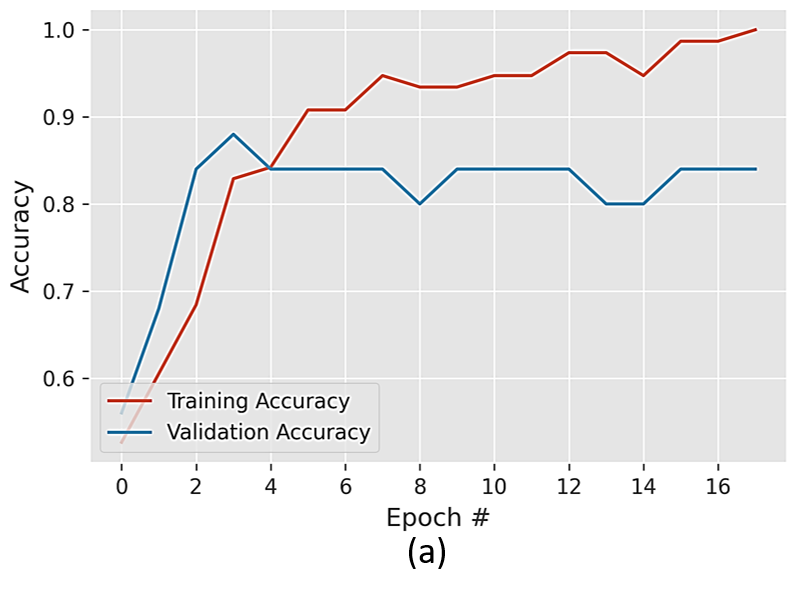}
         \label{fig:results-a}
     \end{subfigure}
     \hfill
     \begin{subfigure}[b]{0.31\textwidth}
         \centering
         \includegraphics[width=\textwidth]{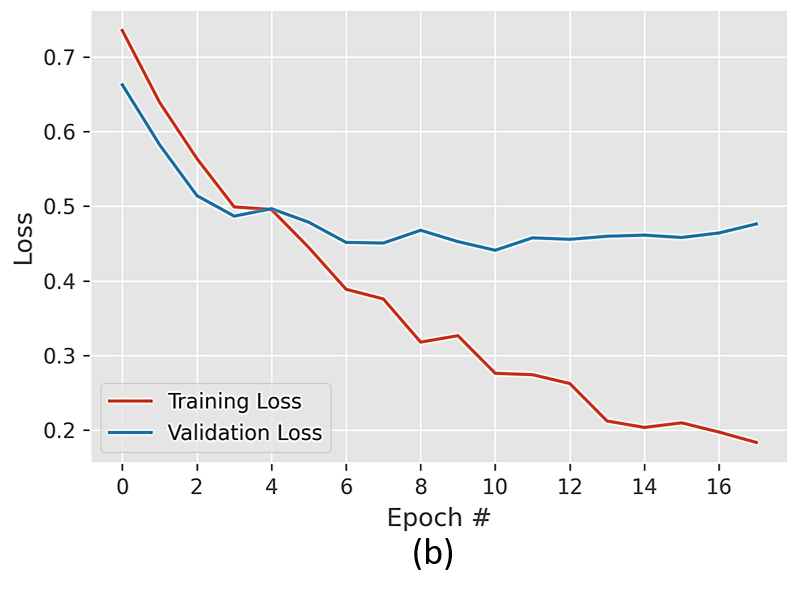}
         \label{fig:results-b}
     \end{subfigure}
     \hfill
     \begin{subfigure}[b]{0.31\textwidth}
         \centering
         \includegraphics[width=\textwidth]{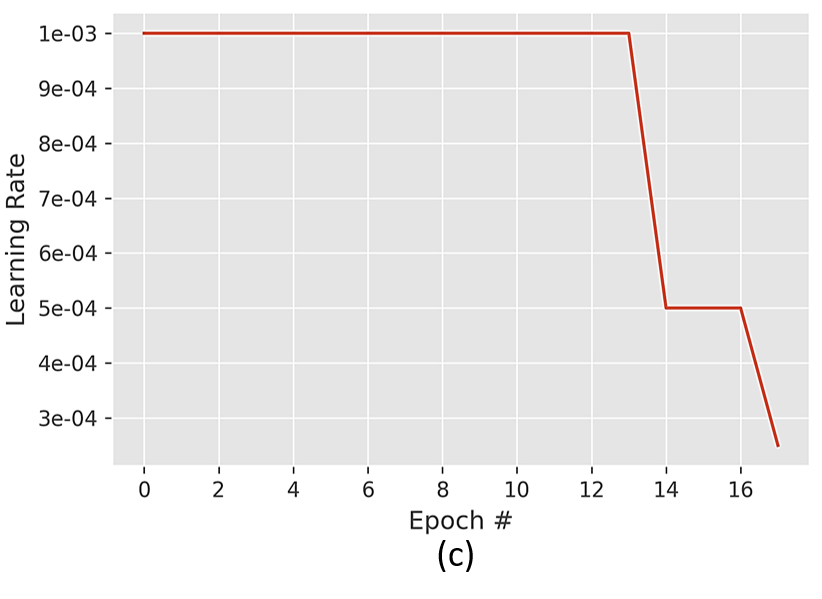}
         \label{fig:results-c}
     \end{subfigure}
        \caption{Learning curves through the process of training and optimizing the network. (a) Accuracy, (b) Loss and (c) Learning rate}
        \label{fig:results}
\end{figure}

The learning curves through the process of training and optimizing the network are illustrated in \cref{fig:results}. The training of the network was stopped right before the loss on the validation set started increasing and avoided overfitting, \cref{fig:results}-b. Following the \textit{learning rate scheduler} strategy, during the process of training, the learning rate was decayed two times through epochs of 15 and 18.
The network learned to classify two classes with an overall accuracy of 94.44\% among all the classes. We received a sensitivity of 100\% and 87.50\% for positive and negative classes, respectively, meaning no false-negative detection for COVID-19 cases. The network achieved a precision of 90.91\% for the positive class and 100\% for the negative class.

\section{Discussion and Future Work}
In this work, we design COVID-Net UV, a hybrid end-to-end network architecture to classify lung POCUS videos for the diagnosis
of COVID-19. Our network comprises two modules: pre-trained InceptionV3 to extract spatial features from video frames and  RNN structure containing GRU units to learn the temporal dependence between video frames. Our results with no false-negative cases for COVID-19 indicate COVID-Net UV is a powerful tool that exceeds human experts with a sensitivity of 86.4\% \cite{Islam2021Thoracic} and models with purely spatial architecture (with highest accuracy of 83.2\%) \cite{Awasthi2021Mini-COVIDNet:}.

Admittedly, we are aware that our proposed network architecture is evaluated on a small dataset containing 119 POCUS videos. However, we believe our proposed methodology and results are the desired baseline for our future work in examining more complex models on the larger POCUS video dataset. The network architecture can be easily retrained by adding new disease categories or feeding new data and be employed for different classification purposes. 

\bibliographystyle{ieeetr}
\bibliography{references.bib}

\begin{thebibliography}{10}

\bibitem{MacLean2021COVIDNet}
A.~MacLean, S.~Abbasi, A.~Ebadi, A.~Zhao, M.~Pavlova, H.~Gunraj, P.~Xi,
  S.~Kohli, and A.~Wong, ``Covid-net us: A tailored, highly efficient,
  self-attention deep convolutional neural network design for detection of
  covid-19 patient cases from point-of-care ultrasound imaging,'' {\em
  arXiv:2108.03131 [cs, eess]}, 8 2021.
\newblock arXiv: 2108.03131.

\bibitem{Amatya2018Diagnostic}
Y.~Amatya, J.~Rupp, F.~M. Russell, J.~Saunders, B.~Bales, and D.~R. House,
  ``Diagnostic use of lung ultrasound compared to chest radiograph for
  suspected pneumonia in a resource-limited setting,'' {\em International
  Journal of Emergency Medicine}, vol.~11, p.~8, 3 2018.
\newblock PMID: 29527652 PMCID: PMC5845910.

\bibitem{Liu2019Deep}
S.~Liu, Y.~Wang, X.~Yang, B.~Lei, L.~Liu, S.~X. Li, D.~Ni, and T.~Wang, ``Deep
  learning in medical ultrasound analysis: A review,'' {\em Engineering},
  vol.~5, pp.~261--275, 4 2019.

\bibitem{Ng2020Imaging}
M.-Y. Ng, E.~Y.~P. Lee, J.~Yang, F.~Yang, X.~Li, H.~Wang, M.~M.-s. Lui,
  C.~S.-Y. Lo, B.~Leung, P.-L. Khong, C.~K.-M. Hui, K.-y. Yuen, and M.~D. Kuo,
  ``Imaging profile of the covid-19 infection: Radiologic findings and
  literature review,'' {\em Radiology: Cardiothoracic Imaging}, vol.~2,
  p.~e200034, 2 2020.
\newblock publisher: Radiological Society of North America.

\bibitem{Roy2020Deep}
S.~Roy, W.~Menapace, S.~Oei, B.~Luijten, E.~Fini, C.~Saltori, I.~Huijben,
  N.~Chennakeshava, F.~Mento, A.~Sentelli, E.~Peschiera, R.~Trevisan,
  G.~Maschietto, E.~Torri, R.~Inchingolo, A.~Smargiassi, G.~Soldati, P.~Rota,
  A.~Passerini, R.~J.~G. van Sloun, E.~Ricci, and L.~Demi, ``Deep learning for
  classification and localization of covid-19 markers in point-of-care lung
  ultrasound,'' {\em IEEE Transactions on Medical Imaging}, vol.~39,
  pp.~2676--2687, 8 2020.
\newblock event: IEEE Transactions on Medical Imaging.

\bibitem{Erfanian2021Automated}
S.~Erfanian~Ebadi, D.~Krishnaswamy, S.~E.~S. Bolouri, D.~Zonoobi, R.~Greiner,
  N.~Meuser-Herr, J.~L. Jaremko, J.~Kapur, M.~Noga, and K.~Punithakumar,
  ``Automated detection of pneumonia in lung ultrasound using deep video
  classification for covid-19,'' {\em Informatics in Medicine Unlocked},
  vol.~25, p.~100687, 2021.
\newblock PMID: 34368420 PMCID: PMC8332742.

\bibitem{Ebadi2021COVIDx-US}
A.~Ebadi, P.~Xi, A.~MacLean, S.~Tremblay, S.~Kohli, and A.~Wong, ``Covidx-us --
  an open-access benchmark dataset of ultrasound imaging data for ai-driven
  covid-19 analytics,'' {\em arXiv:2103.10003 [cs, eess]}, 4 2021.
\newblock arXiv: 2103.10003.

\bibitem{Azimi2022Improving}
H.~Azimi, J.~Zhang, P.~Xi, H.~Asad, A.~Ebadi, S.~Tremblay, and A.~Wong,
  ``Improving classification model performance on chest x-rays through lung
  segmentation,'' {\em arXiv:2202.10971 [cs, eess]}, 2 2022.
\newblock arXiv: 2202.10971.

\bibitem{Islam2021Thoracic}
N.~Islam, S.~Ebrahimzadeh, J.-P. Salameh, S.~Kazi, N.~Fabiano, L.~Treanor,
  M.~Absi, Z.~Hallgrimson, M.~M. Leeflang, L.~Hooft, v.~d. C.~B. Pol,
  R.~Prager, S.~S. Hare, C.~Dennie, R.~Spijker, J.~J. Deeks, J.~Dinnes,
  K.~Jenniskens, D.~A. Korevaar, J.~F. Cohen, d.~A.~V. Bruel, Y.~Takwoingi,
  v.~d.~J. Wijgert, J.~A. Damen, J.~Wang, M.~D. McInnes, and C.~C.-. D. T.~A.
  Group, ``Thoracic imaging tests for the diagnosis of covid‐19,'' {\em
  Cochrane Database of Systematic Reviews}, vol.~2021, no.~3, 2021.
\newblock [Online; accessed 2022-03-04].

\bibitem{Awasthi2021Mini-COVIDNet:}
N.~Awasthi, A.~Dayal, L.~R. Cenkeramaddi, and P.~K. Yalavarthy,
  ``Mini-covidnet: Efficient lightweight deep neural network for ultrasound
  based point-of-care detection of covid-19,'' {\em IEEE Transactions on
  Ultrasonics, Ferroelectrics, and Frequency Control}, vol.~68, pp.~2023--2037,
  6 2021.
\newblock event: IEEE Transactions on Ultrasonics, Ferroelectrics, and
  Frequency Control.

\end{thebibliography}

\end{document}